\documentclass[aps,prl,twocolumn,showpacs,superscriptaddress,groupedaddress]{revtex4}
\usepackage{amssymb}
\usepackage{amsmath}
\usepackage{graphicx}
\usepackage{epstopdf}
\usepackage{subfigure}
\usepackage{natbib}
\usepackage{epsfig}
\usepackage{amsfonts}
\usepackage{mathrsfs}
\usepackage[toc,page,title,titletoc,header]{appendix}
\usepackage[colorlinks,linkcolor=blue,citecolor=blue,anchorcolor=blue]{hyperref}
\usepackage{dsfont,amsthm,amsbsy}
\def\be{\begin{equation}}
\def\ee{\end{equation}}

\def\bi{\begin{itemize}}
\def\ei{\end{itemize}}
\def\bn{\begin{enumerate}}
\def\en{\end{enumerate}}
\def\bea{\begin{eqnarray}}
\def\eea{\end{eqnarray}}
\def\no{\nonumber}
\def\ba{\begin{array}}
\def\ea{\end{array}}
\def\bd{\begin{displaymath}}
\def\ed{\end{displaymath}}

\begin{document}

\title{Loschmidt Echo Revivals: Critical and Noncritical}

\author{R. Jafari}
%\email{Corresponding author: rouhollah.jafari@physics.gu.se, rohollah.jafari@gmail.com}
\affiliation{Beijing Computational Science Research Center, Beijing 100094, China}
\affiliation{Department of Physics, University of Gothenburg, SE 412 96 Gothenburg, Sweden}
\affiliation{Department of Physics, Institute for Advanced
Studies in Basic Sciences (IASBS), Zanjan 45137-66731, Iran}
\author{Henrik Johannesson}
%\email[]{henrik.johannesson@physics.gu.se}
\affiliation{Beijing Computational Science Research Center, Beijing 100094, China}
\affiliation{Department of Physics, University of Gothenburg, SE 412 96 Gothenburg, Sweden}

%\date{\today}
\begin{abstract}

A quantum phase transition is generally thought to imprint distinctive
characteristics on the nonequilibrium dynamics of a closed quantum system. Specifically,
the Loschmidt echo after a sudden quench to a quantum critical point $-$ measuring the
time dependence of the overlap between initial and time-evolved states $-$ is expected to
exhibit an accelerated relaxation followed by periodic revivals.
We here introduce a new exactly solvable model, the extended Su-Schrieffer-Heeger model,
the Loschmidt echo of which provides a counterexample. A parallell
analysis of the quench dynamics of the three-site spin-interacting $XY$ model allows us to pinpoint
the conditions under which a periodic Loschmidt revival actually appears.

%makes us conclude that quantum criticality is neither a sufficient nor a necessary condition for periodic Loschmidt echo revivals.

\end{abstract}


\maketitle

Taking a quantum system out of equilibrium can be done in many ways, such as injecting energy through an external reservoir or
applying a driving field. The simplest paradigm is maybe that of a {\em quantum quench}, where a closed system is pushed out of equilibrium by a sudden change in the Hamiltonian which controls its time evolution. Studies of quantum quenches have spawned a large body of results, on equilibration and thermalization \cite{Gogolin2016} (and its breakdown in integrable systems \cite{Vidmar2016}), on entanglement dynamics \cite{Calabrese2016}, and more \cite{Polkovnikov2011,Mitra2016}. In this context, an important task is to identify nonequilibrium dynamical signatures of a quantum phase transition (QPT). The problem comes in a variety of shapes, ranging from the Kibble-Zurek mechanism for defect production \cite{Zurek2005} to the time evolution of correlations in strongly correlated out-of-equilibrium systems at a QPT \cite{Manmana2009}. A basic variant is to ask the question: If a Hamiltonian is suddenly quenched to a quantum critical point (or its vicinity), is there any special characteristic of the subsequent dynamics?

To address this question one may invoke the {\em Loschmidt echo} (LE) \cite{Gorin}, which measures the overlap between the initial (prequench) and time-evolved (postquench) state. %The concept of the LE has a long history which goes beyond the confines of quantum quenches, and $-$ being a measurable quantity $-$ has appeared in a multitude of problems %dealing with time evolution of quantum systems.
Applied to a {\em quantum critical quench} $-$ i.e. with the quench parameter pulled to a quantum critical point $-$ finite-size case studies reveal that the time dependence of the LE of several models exhibits a periodic pattern, a {\em revival structure}, formed by brief detachments from its mean value \cite{Quan, Yuan2007, Rossini2007-a, Zhong2011, Happola, Montes, Sharma2012, Rajah2014}, implying revivals also for expectation values of local observables \cite{Igloi2011, Cardy2014}. The amplitudes of these revivals may decay with time, however, their presence appears to be independent of the initial state and the size of the quench \cite{Happola}. Indeed, the distinctive structure of revivals of the LE after a quench has been conjectured to be a faithful witness of quantum criticality \cite{Quan, Yuan2007}.

In this Letter we challenge the notion that quantum criticality and LE revival structures are intrinsically linked. We do this by way of example, introducing a new exactly solvable model, the {\em extended Su-Schrieffer-Heeger} (ESSH) {\em model}, which exhibits several distinct quantum phases with associated QPTs. The ESSH model serves as a representative of a large class of quasifree 1D Fermi systems, and contains as special cases the original SSH model \cite{Su},  the Creutz model \cite{Creutz}, and the Kitaev chain \cite{Kitaev2001} and its dimerized version \cite{Wakatsuki2014}. Moreover, via a Jordan-Wigner transformation \cite{Jordan}, and with suitably chosen parameters, the ESSH model embodies several generic spin chain models, including the 1D quantum compass model \cite{Nussinov2015}. Important for the present work, the quench dynamics of the ESSH model highlights the conditions under which the LE may show a revival structure. Informed by this, and by results extracted from another exactly solvable model, the {\em three-site spin-interacting (TSSI) $XY$ model} \cite{Titvinidze, Zvyagin}, we come to the conclusion that quantum criticality is neither a sufficient nor a necessary condition for the LE to exhibit an observable revival structure. Instead, what matters is that the quasiparticle modes which control the LE are massless {\em {and}} have a group velocity $v_g \!\gg \!L/t$, where $L$ is the length of the system and $t$ is the observation time. Only if these modes coincide with the quantum critical modes is a revival structure tied to a QPT. These conditions, which are general, bring
new light on the important issue of how to read a LE after a quantum quench.

\indent {\em Loschmidt echo. $-$} \ A quantum quench is a sudden change in the Hamiltonian
$H(\theta_1)$ of a quantum system, with $\theta_1$ denoting the value(s)
of the parameter(s) that will be quenched. The system is initially prepared in an eigenstate
$|\Psi_{m}(\theta_1)\rangle$ to the Hamiltonian $H(\theta_1)$. The quench is carried out at time $t=0$,
when $\theta_1$ is suddenly switched to $\theta_2$. The system then evolves with
the quench Hamiltonian $H(\theta_2)$ according to  $|\Psi_{m}(\theta_1, \theta_2,t)\rangle=\exp(-iH(\theta_2)t) |\Psi_{m}(\theta_1)\rangle$.
In this case the LE \cite{Gorin}, here denoted by ${\cal L} (\theta_1,\theta_2,t)$, reduces to a dynamical version of the ground-state fidelity (return probability),
\be
{\cal L} (\theta_1,\theta_2,t)= |\langle\Psi_{m}(\theta_1)|\exp(-iH(\theta_2)t) |\Psi_{m}(\theta_1)\rangle|^2,
\ee
measuring the distance between the time-evolved
state $|\Psi_{m}(\theta_1,\theta_2,t)\rangle$ and the initial state $|\Psi_{m}(\theta_1)\rangle$.
%By considering the ground state of the system as the initial state, the
%LE can be interpreted as a dynamical version of the squared {\em ground-state fidelity}
%$F(\theta,\theta^{\prime})$ \cite{Venuti2010}, defined by the overlap between two ground
%states at different parameter values
%$\theta$ and $\theta^{\prime}$:
%$F(\theta,\theta^{\prime})=|\langle \Psi_{0}(\theta)|\Psi_{0}(\theta^{\prime})\rangle|$.

The LE typically decays
in a short time $T_{\text{rel}}$ {\em (relaxation time)}, from unity to some mean value around which it then
fluctuates \cite{Venuti2011}.  {\em Revivals} are also visible in the LE as pronounced deviations
from the average value %(conventionally defined by a lower bound of three standard deviations
\cite{Happola}.
For quenches to a quantum critical point in a finite system there is an expectation
that the LE relaxation is accelerated \cite{Quan, Yuan2007, Zhang2009, Rossini2007-a, Rossini2007-b, Sharma2012, Sacramento} and that the revivals are periodic \cite{Quan, Yuan2007, Happola, Montes}.
Conversely, such behavior has been proposed as a signature of quantum criticality \cite{Quan, Yuan2007}.
However, the matter turns out to be more complex. To see how, we next introduce the ESSH model
and exhibit its quench dynamics.

{\em Extended Su-Schrieffer-Heeger (ESSH) model.} $-$ \newline \noindent We define the Hamiltonian of the ESSH model by
%\be
%\begin{aligned}
%\label{eq1}
%&
%H =
%\sum_{n=1}^{N}
%\Big[-(wc^{A\dagger}_{n}c^{B}_{n}+\tau c^{A\dagger}_{n+1}c^{B}_{n}+\Delta e^{-i\theta} c^{A\dagger}_{n}c^{B\dagger}_{n}\\
%&\hspace{0.3cm}
%+\Lambda e^{i\theta}c^{A\dagger}_{n+1}c^{B\dagger}_{n}) + \frac{\mu}{2} (c^{A\dagger}_{n}c^A_{n}+c^{B\dagger}_{n}c^B_{n})\Big] +\mbox{H.c}.
%\end{aligned}
%\ee
\begin{eqnarray}
H = \sum_{n=1}^{N}
\Big[&\!\!-\!\!&(wc^{A\dagger}_{n}c^{B}_{n}\!+\!\tau c^{A\dagger}_{n+1}c^{B}_{n}\!+\!\Delta e^{-i\theta} c^{A\dagger}_{n}c^{B\dagger}_{n} \\
&\!\!+\!\!&\Lambda e^{i\theta}c^{A\dagger}_{n+1}c^{B\dagger}_{n}) \!+\! \frac{\mu}{2} (c^{A\dagger}_{n}c^A_{n}\!+\!c^{B\dagger}_{n}c^B_{n})\Big] \!+\!\mbox{H.c.}, \nonumber
\end{eqnarray}
where $A$ and $B$ are sublattice indices labeling fermion creation and annihilation operators $c^{A/B \dagger}_{n}$ and $c^{A/B}_n$, $w$ and $\tau$ are
hopping amplitudes, $\Delta$ and $\Lambda$ are superconducting
pairing gaps, $\pm \theta$ are the phases of the pairing terms, and $\mu$ is a chemical potential.
Choosing $\mu=0$ and introducing the Nambu spinor $\Gamma^{\dagger}=
(c^{A\dagger}_{k},c^{B\dagger}_{k},c^{A}_{-k},c^{B}_{-k})$,
the Fourier transformed Hamiltonian can be expressed in
Bogoliubov-de Gennes (BdG) form \cite{Zhu2016}, $H= \sum_{k\ge0}\Gamma^{\dagger}H(k)\Gamma$, with
\bea
\label{BdG}
%\no
H(k)=
\left(
  \begin{array}{cccc}
    0 & p_{k} & 0 & q_{k} \\
    p_{k}^{\ast} & 0 & -q_{-k} & 0 \\
    0 & -q_{-k}^{\ast} & 0 & -p^{\ast}_{-k} \\
    q_{k}^{\ast} & 0 & -p_{-k} & 0 \\
  \end{array}
\right),
\eea
where $\quad p_{k}\!=\!-(w+\tau e^{-ika})$ and $q_{k}\!=\!-(\Delta e^{-i\theta}-\Lambda e^{i(\theta-ka)})$. Here $k=2m\pi/L$, $m=0,\cdots,N/2$, given periodic boundary conditions, and $L=Na$ with $a$ the lattice spacing, taken as unity in arbitrary units.

By diagonalizing $H(k)$ one obtains the quasiparticle Hamiltonian
$H=\sum_{\alpha=1}^{4}\sum_{k}\varepsilon^{\alpha}_{k}\gamma_{k}^{\alpha\dag}\gamma_{k}^{\alpha}$, with $\gamma_{k}^{\alpha\dag}$ and $\gamma_{k}^{\alpha}$ linear combinations of the elements in the Nambu spinor, and with corresponding energy bands
$\varepsilon^{1}_{k}=-\varepsilon^{4}_{k}=-\sqrt{a_k+\sqrt{a_k^{2}-b_k}}$ and $\varepsilon^{2}_{k}=-\varepsilon^{3}_{k}=-\sqrt{a_k-\sqrt{a_k^{2}-b_k}}$,
where $a_k=|q_{k}|^{2}+|p_{k}|^{2}+|q_{-k}|^{2}+|p_{-k}|^{2}$ and
$b_k=4(p_{k}^{\ast}p_{-k}-q_{k}^{\ast}q_{-k})(p_{k}p_{-k}^{\ast}-q_{k}q_{-k}^{\ast})$. The ground state $|\Psi_0\rangle$ is obtained by filling up the negative-energy quasiparticle states,
$|\Psi_0\rangle = \prod_k \gamma_k^{2 \dag} \gamma_k^{1 \dag} |V\rangle$, where $|V\rangle$ is the Bogoliubov vacuum annihilated by the $\gamma_k$'s (see Supplemental Material \cite{Jafari2016}).

One easily verifies that the gap to the first excited state vanishes for all momenta $k$ when $\theta\! =\!\pi/2$,
$w=\Delta$, and $\tau=\Lambda$. The ground state here acquires a degeneracy of
$2^{N/2}$ (enlarged to $2\times2^{N/2}$ at the isotropic point (IP) $\Delta=\Lambda$) \cite{Jafari2016}.
%at $k\!=\!0$ when
%$\theta \!=\! \arccos[(\Delta^{2}+\Lambda^{2}-(w+\tau)^{2})/2\Delta\Lambda]/2$,
%and at $k\!=\! \pi$ when
%$\theta \!= \! \arccos[((w-\tau)^{2}-\Delta^{2}-\Lambda^{2})/2\Delta\Lambda]/2$.
It follows that the line $\theta\!=\!\pi/2$ in parameter space is critical for any
ratio $\Delta/\Lambda$. Its interpretation is most easily phrased in spin language by connecting
the ESSH model to the general
quantum compass model \cite{You2014, Nussinov2015} via a Jordan-Wigner transformation \cite{Jordan}.
The critical line $\theta\!=\!\pi/2$ is then seen to define a (nontopological) QPT between two distinct phases
with large short-range spin correlations in the $x$ and
$y$ direction respectively.  As expected \cite{Zanardi2007}, this QPT is signaled by a sharp decay
of the ground-state fidelity $F(\theta,\theta \!+\! \delta \theta) = |\langle \Psi_{0}(\theta)|\Psi_{0}(\theta^{\prime})\rangle|$,
cf. Fig. \ref{figS2} in \cite{Jafari2016}.

\begin{figure}[h]
\centerline{
\includegraphics[width=0.60\linewidth,height=0.40\linewidth]{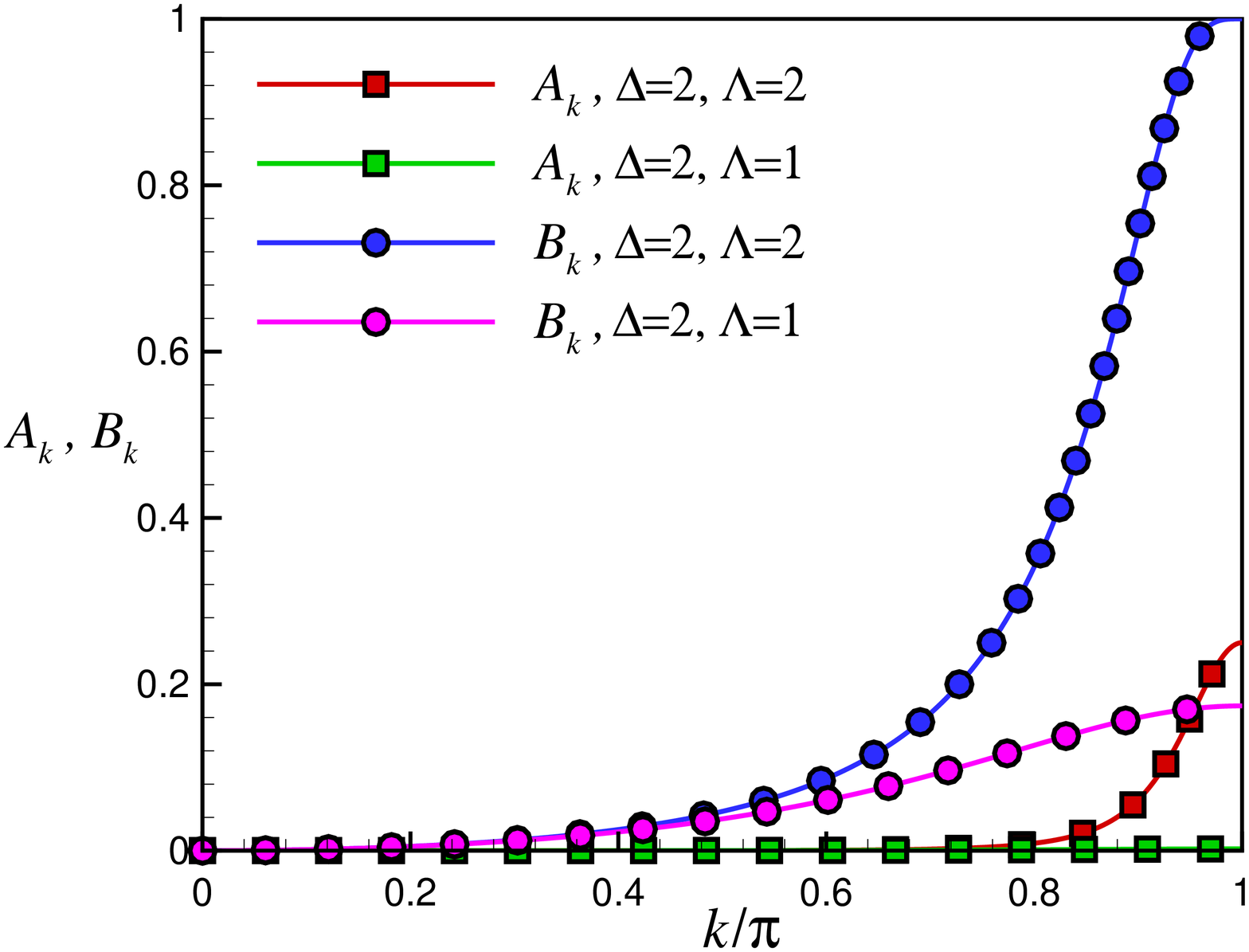}}
\caption{(Color online) The amplitudes $A_{k}$ and $B_{k}$ in Eq. (\ref{Loschmidtmodes}) plotted versus $k$
at the isotropic point $w=\Delta=\tau=\Lambda=1$ and away from the isotropic point
 $w=\Delta=2$, $\tau=\Lambda=1$.\\ }
\label{figA}
\end{figure}

{\em Loschmidt echo in the ESSH model.} $-$
By a rather lengthy calculation one can obtain the complete set of eigenstates of the model,
yielding an exact expression for the LE \cite{Jafari2016}
When the system is initialized in the ground state
$|\Psi_0(\theta_1)\rangle$ and quenched to the critical line, i.e. with
$\theta_2=\theta_{c}=\pi/2$, one obtains
\begin{multline}
\label{Loschmidtmodes}
{\cal L}({\theta_1},{\theta_c},t) \\ =
\prod_{0 \le k \le \pi}|1\!-\!A_k\sin^2(\varepsilon_{k}^{1}(\theta_{c})t)\!-\!B_k\sin^2(\frac{\varepsilon_{k}^{1}(\theta_{c})t}{2})|,
\end{multline}
where $A_{k}$ and $B_{k}$ measure overlaps between $k$ modes of the initial ground state, $|\psi_{0,k}(\theta_1)\rangle$, and
eigenstates $|\psi_{m,k}(\theta_c)\rangle$ of $H(\theta_c)$; cf. Fig. \ref{figA} and \cite{Jafari2016}. The energies $\varepsilon_{k}^{1}(\theta_{c})$
are those of the quasiparticles in the lowest filled band in the ground state of the critical quench Hamiltonian.
%%%%%%%%%%%%%%%%%%%%%%%%  Fig.S2   %%%%%%%%%%%%%%%%%%%%%%%
%%%%%%%%%%%%%%%%%%%%%%%%%%%%%%%%%%%%%%%%%%%%%%%%%%%%%%%

\begin{figure}[h]
%\begin{minipage}{0.95\linewidth}
%\small{(a)} $\delta$
%\includegraphics[width=0.7\linewidth]{./figures/fig3.eps}
%\centering
%\end{minipage}
\begin{minipage}{.95\linewidth}
%\small{(b)} $\delta$
\includegraphics[width=0.85\linewidth,height=0.7\linewidth]{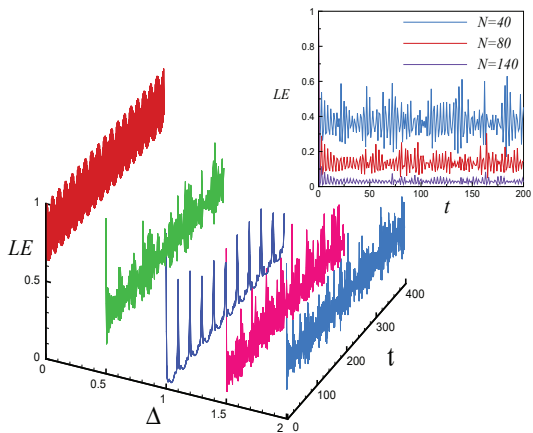}
\centering
\end{minipage}
\caption{ (Color online)
%The LE as a function of $\theta_2$ and time $t$, where
%the Hamiltonian parameters are chosen as $\Delta=\Lambda = \tau = w=1$, for
%$N=80$ and with $\delta \theta= \theta_2-\theta_1=0.001$ the magnitude of the quantum quench.
The LE versus $\Delta$ and time $t$ for quenches to the critical line
$\theta_2=\theta_c=\pi/2$ starting from $\theta_1=0.45\pi$, for $w=\Delta$, $\tau=\Lambda=1$
and $N=40$.
Inset: The LE versus time $t$ for quenches to the critical line
$\theta_c=\pi/2$ starting from $\theta_1=0.45\pi$, for different system sizes $N$ and
with $w=\Delta=2$, $\tau=\Lambda=1$.}
\label{fig2}
\end{figure}

%In Fig. \ref{fig2}(a) we have plotted ${\cal L}({\theta_1},{\theta_2},t)$ as a function of $\theta_2$ and $t$ at the IP for $N=80$, with the small
%quench $\delta \theta = \theta_2 - \theta_1= 0.001$ making it the dynamical counterpart of the fidelity in Fig. (S2) in \cite{Jafari2016}.
%One clearly sees a rapid decay of the
%LE when quenching to the critical line $\theta_c=\pi/2$, with periodic revivals in time. This is in agreement with several studies of LEs
%at quantum criticality \cite{Quan, Yuan2007, Venuti2011, Happola, Montes, Zhang2009, Haikka2012, Rajah2014, Rossini2007-a, Rossini2007-b, Sharma2012, Zhong2011}.
%However, departing from the IP, taking $\Delta\neq\Lambda$, but remaining at the critical line $\theta_c = \pi/2$,
%a surprising result occurs: The periodic revivals get wiped out for sufficiently large anisotropies,
%with the LE oscillating randomly around its
%mean value; cf. Fig. \ref{fig2}(b) for $\delta \theta = 0.05 \pi$.
In Fig. \ref{fig2} we have plotted ${\cal L}({\theta_1},{\theta_2},t)$ versus $\Delta$ and time $t$ for quenches to the critical line
$\theta_2=\theta_c=\pi/2$ starting from $\theta_1=0.45\pi$, for $w=\Delta$, $\tau=\Lambda=1$ and $N=40$. One clearly sees a rapid decay of the
LE, with periodic revivals in time when quenching to the IP $\Delta=1$. This is in agreement with several studies of LEs
at quantum criticality \cite{Quan, Yuan2007, Venuti2011, Happola, Montes, Zhang2009, Haikka2012, Rajah2014, Rossini2007-a, Rossini2007-b, Sharma2012, Zhong2011}.
However, departing from the IP, taking $\Delta\neq\Lambda$, but remaining at the critical line $\theta_c = \pi/2$,
a surprising result occurs: The periodic revivals get wiped out
for sufficiently large anisotropies,
with the LE oscillating randomly around its mean value.
%%%%%%%%%%%%%%%%%%%%%%%  Fig.2   %%%%%%%%%%%%%%%%%%%%%%%
%%%%%%%%%%%%%%%%%%%%%%%%%%%%%%%%%%%%%%%%%%%%%%%%%%%%%%%
%%%%%%%%%%%%%%%%%%%%%%%  Fig.3   %%%%%%%%%%%%%%%%%%%%%%%
\begin{figure}[h]
\begin{minipage}{0.85\linewidth}
%\small{(a)} $\delta$
\includegraphics[width=0.8\linewidth,height=0.55\linewidth]{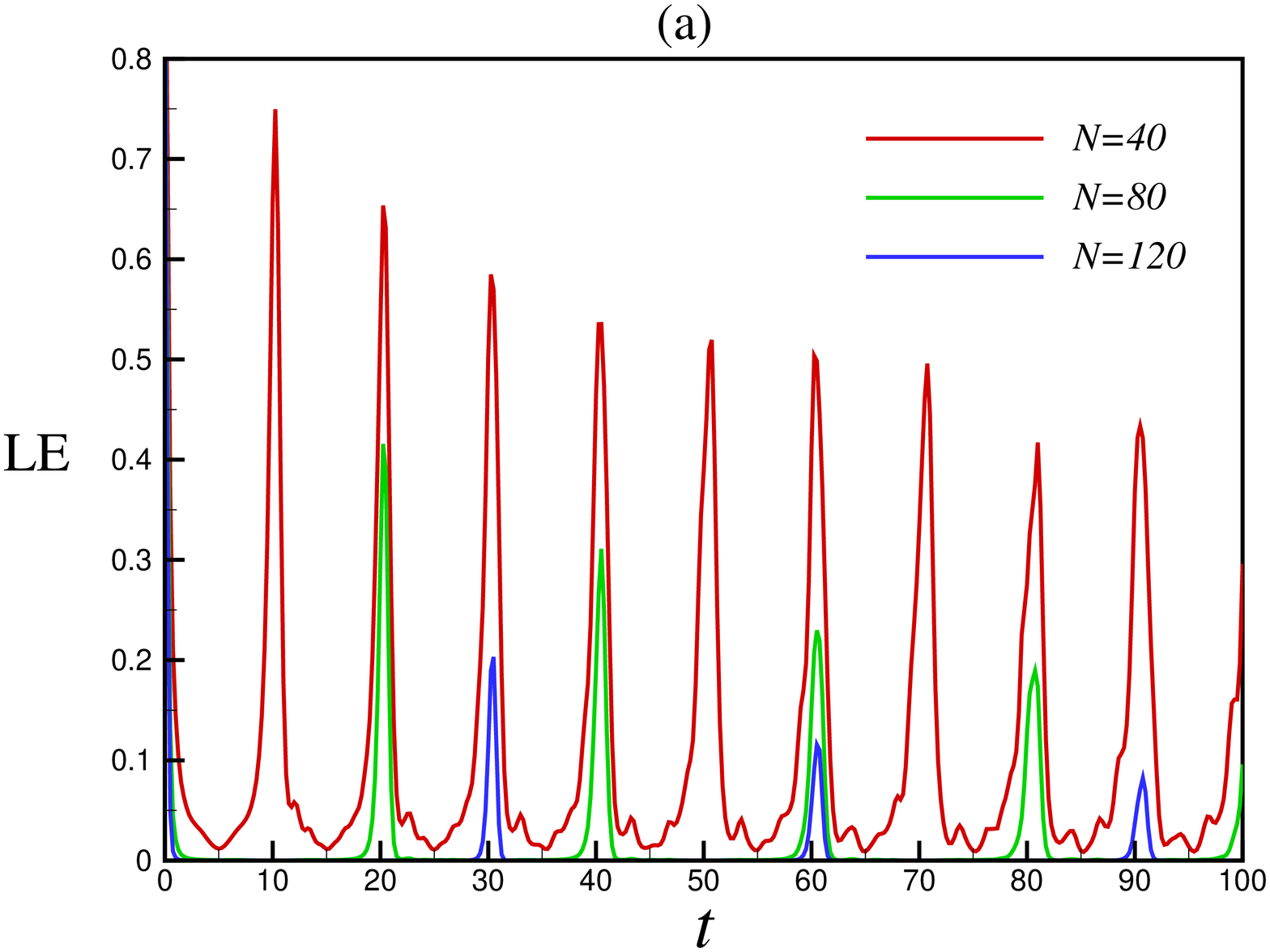}
\centering
\end{minipage}
\begin{minipage}{0.85\linewidth}
%\small{(b)} $\delta$
\includegraphics[width=0.8\linewidth,height=0.55\linewidth]{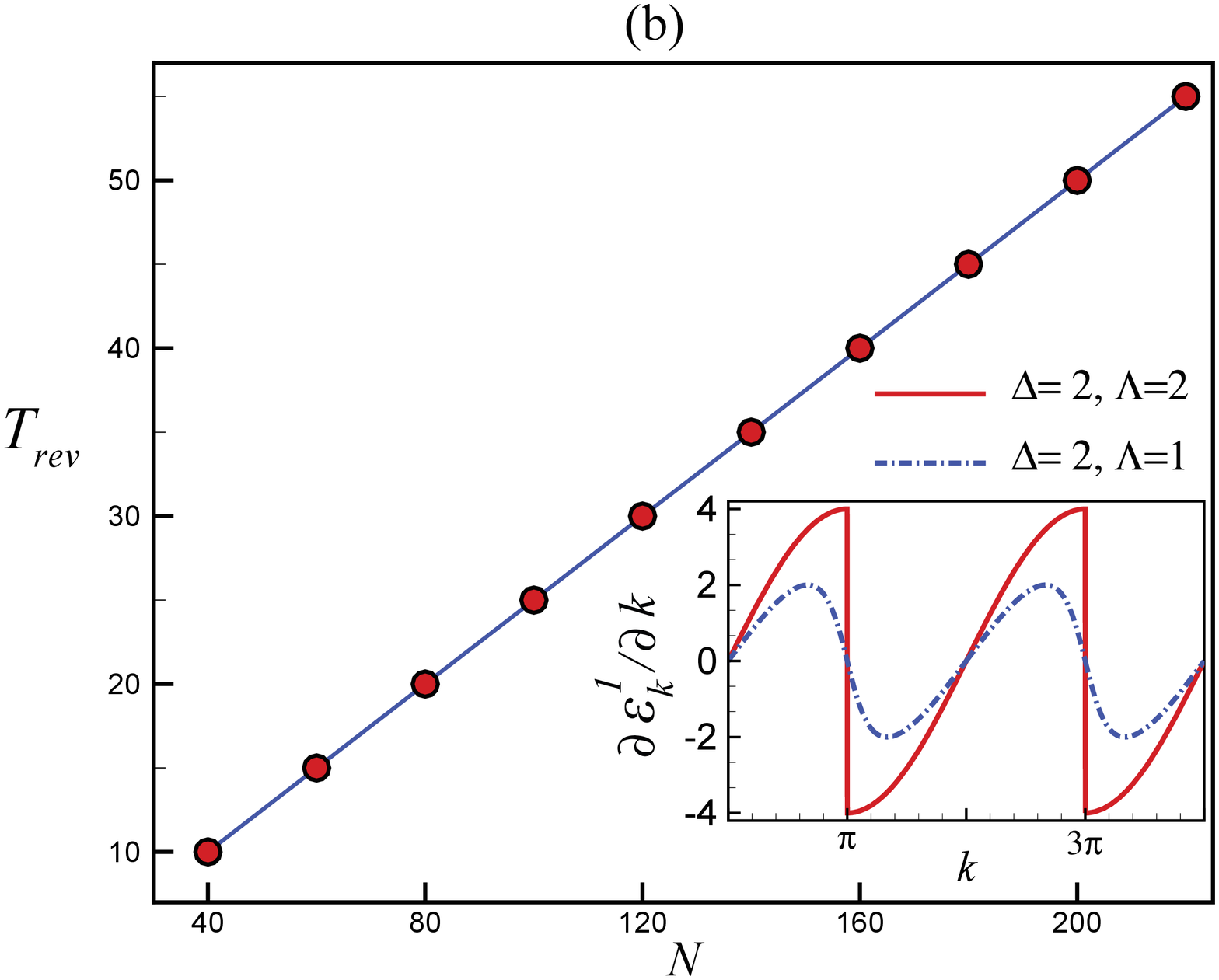}
\centering
\end{minipage}
\caption{(Color online) (a) LE versus
time $t$, with initial pairing phase $\theta_1=0.45\pi$ and quenching to the critical line $\theta_c=\pi/2$,
for various system sizes $N$ at the IP $w=\Delta=2$, $\tau=\Lambda=2$.
(b) Scaling of the revival period $T_{\text{rev}}$
with system size $N$ for a quench to the critical line at the IP.
Inset: The derivative of the ground-state energy modes $\epsilon_k^1$ {\em (group velocity)}
at the critical line $\theta = \pi/2$ for
isotropic (red line) and anisotropic (blue hatched line) cases.}
%Inset (top): Linear scaling of the time at which the LE
%reaches its first minimum with system size for a quench to the critical point
%for isotropic case. %$w=\Delta=2$, $\tau=\Lambda=2$ with periodic boundary condition.
\label{fig3}
\end{figure}
%%%%%%%%%%%%%%%%%%%%%%%%%%%%%%%%%%%%%%%%%%%%%%%%%%%%%%%

To find out why the LE exhibits a revival structure at or very close to the IP, but not farther
away from the IP,
let us begin by pinpointing the revival periods at the IP, manifest in
Fig. \ref{fig3}(a).
Plotting $T_{\text{rev}}$ versus $N$, cf. \!Fig. \ref{fig3}(b), unveils a linear scaling
\bea
\label{scaling}
T_{\text{rev}}=\frac{Na}{K},
%T_{\text{rev}}=KN^{\eta},d
\eea
where $K$ has dimension of velocity with value $K\! =\! 4.00 \!\pm 0.03$.
A numerical spectral analysis suggests that $K\! \approx v_{\text{max}}$, where $v_{\text{max}} \!= \!\mbox{max}(\partial_k \varepsilon^1_{k}(\theta_c))$, cf. inset, Fig. \ref{fig3}(b).
This result is anticipated from a study of the spin-1/2 $XY$ model \cite{Happola}, where the LE revival period is also governed by the maximum quasiparticle
group velocity produced by the critical quench Hamiltonian. However, Eq. (\ref{scaling}), with $K \approx v_{\text{max}}$, fails to account for the disappearance of periodic
revivals away from the IP. Why is that?

The answer lies in Eq. (\ref{Loschmidtmodes}). First note that a revival requires that all $k$ modes in Eq. (\ref{Loschmidtmodes})
contribute sizably to the LE, in turn requiring that the oscillating terms are small. An analysis shows that the oscillation amplitudes $A_k$ and $B_k$
are indeed small except for $B_k$ when approaching the BZ boundary (at which $B_k$ takes its maximum), cf. Fig. \ref{figA}.
It follows
that the corresponding modes can contribute constructively to the LE only at time instances at which their oscillation terms get suppressed.
Thus, we expect that the most pronounced revivals happen when the vanishing of the term
proportional to $B_{k=\pi}$ is concurrent with the near vanishing of $B_k$ terms with $k$ close to $\pi$.
To obtain the revival period at the IP we thus make the ansatz $\varepsilon_{k_0}^{1}(\theta_{c})\,t/2=m\pi$, with $m$ an integer and with $k_0$
the mode with the largest group velocity in the vicinity of the BZ boundary.
A Taylor expansion to first order, $\varepsilon_{k_0-p\delta k}^{1}(\theta_{c})
\approx \varepsilon_{k_0}^{1}(\theta_{c})\!-\!\partial_k \epsilon_{k}^{1}(\theta_{c})|_{k_0}\, p\delta k$
shows that $B_k$ terms of neighboring $k$ modes are strongly suppressed whenever
$t$ is a multiple of $Na/v_{\text{max}}$ with $v_{\text{max}}=\partial_k \epsilon_{k}^{1}(\theta_{c})|_{k_0}$ and (as before) $a=1$. Here $p\ll N$ are integers
and $\delta k = 2\pi/N$. This estimate of the revival period agrees with the numerical result in Eq. (\ref{scaling}).

Turning to the anisotropic case $\Delta \neq \Lambda$ and repeating the analysis from above immediately reveals why the revival structure now gets lost.
First, as exemplified in Fig. \ref{figA}, the $B_k$ amplitudes are here small for {\em all} $k$ modes. Thus, the simultaneous suppression of the dominant (but still small) oscillation terms is not expected to have a significant effect on the LE. Moreover, as seen in the inset of Fig. \ref{fig3}(b), the group velocities $v_k = \partial_k \epsilon_{k}^{1}(\theta_{c})$ away from the IP are quite small throughout the $k$ range where $B_k$ is nonvanishing.  As a consequence, with $T_{\text{rev}} \approx L/v_{k=\pi}$ (as before obtained by expanding the quasiparticle energies close to $k=\pi$ where the $B_k$ amplitudes are largest),
one would have to wait an exceedingly long time to see any trace of a weak revival structure, if at all present.

To understand the origin of the different behaviors of the LE at the IP and away from the IP, recall from Eq. (\ref{Loschmidtmodes}) that the revivals are controlled by quasiparticles in the lowest energy band, $\varepsilon_k^1$. This is so since the second filled quasiparticle band in the ground state, $\varepsilon_k^2$, collapses to zero and becomes dispersionless at the critical line $\theta_c=\pi/2$ \cite{Jafari2016}. Away from the IP, the $\varepsilon_k^1$ band remains gapped for all $k$ also at the critical line, thus holding back quasiparticle excitations from that band. This is different from the critical line at the IP where the gap closes at the BZ boundary \cite{Jafari2016}. Since the oscillation amplitudes can be interpreted as measuring the probabilities of quasiparticle excitations, $k$ modes at or near the gap-closing points are indeed expected to yield much larger amplitudes. As follows from our result for the revival period, if these modes also give rise to a group velocity $v_g \gg L/t$, with $t$ the observation time, a revival structure will ensue. Note that here $v_g$ is the group velocity of quasiparticles at which the oscillation amplitudes peak. While $v_g$ happens to be at a global maximum in the ESSH model at the IP, this property is not expected to be generic.

{\em Loschmidt echo in the three-site spin-interacting $XY$ model.} $-$ Having established that quantum criticality is not a sufficient condition for a revival structure in a LE, what about the converse? Can a LE exhibit a revival structure without the presence of a QPT?

The answer is yes. A case in point is the LE of a quench to the $h_s=0$ line in the $J_3$-$h$ parameter space of the three-site spin-interacting (TSSI) XY model \cite{Titvinidze, Zvyagin},
\begin{eqnarray}
\label{TSSI}
H_{\text{TSSI}} =  \!&-&\frac{J}{2}\sum_{j=1}^N (\sigma_j^x \sigma^x_{j+1} \!+\! \sigma_j^y \sigma^y_{j+1}) - h_s\sum_{j=1}^N (-1)^j\sigma^z_j \nonumber \\
&-&\frac{J_3}{4} \sum_{j=1}^N (\sigma_j^x \sigma^x_{j+2} + \sigma_j^y \sigma^y_{j+2})\,\sigma^z_{j+1}
\end{eqnarray}
where $\sigma^x, \sigma^y$, and $\sigma^z$ are the usual Pauli matrices. In Ref. \cite{Divakaran} it was noted that the decay rate of the LE shows an accelerated decay in such a quench, independent of whether the quench is critical ($J_3 \!=\! 0$) or noncritical ($J_3 \!\neq \!0$).  In contrast, the LEs of quenches to the critical lines $h_s \!=\! \pm J_3/2$ which define a QPT between an antiferromagnetic and type-I spin-liquid phase display neither enhanced decays nor revival structures. %The author of Ref. \cite{Divakaran} argued that this %anomalous behavior comes about because the reduced effective Hamiltonian matrix depends on the three-spin term ($\sim J_3$) only via its diagonal elements. However, the LE in the %{\em cluster XY model} does show an accelerated decay and revival structure after a critical quench in the presence of a three-spin interaction, \cite{Montes} casting doubt on this %explanation.
%
%
%%%%%%%%%%%%%%%%%%%%%%%  Fig.4   %%%%%%%%%%%%%%%%%%%%%%%
\begin{figure}
\begin{minipage}{0.95\linewidth}
%\small{(a)} $\delta$
\includegraphics[width=0.7\linewidth,height=0.5\linewidth]{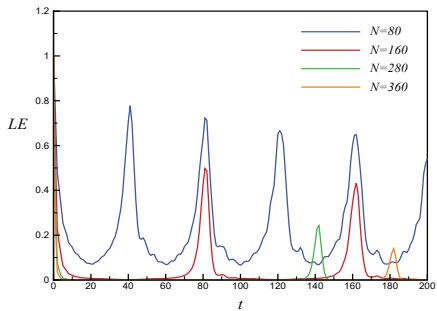}
\centering
\end{minipage}
\caption{(Color online) The LE of the TSSI $XY$ model
versus time $t$ at the noncritical point where $J_3=4$ and $h_s=0$.}
\label{fig4}
\end{figure}
%%%%%%%%%%%%%%%%%%%%%%%%%%%%%%%%%%%%%%%%%%%%%%%%%%%%%%%
%
%

Guided by our results for the ESSH model, we resolve this conundrum by numerically confirming that the absence of a revival structure for a quench from the antiferromagnetic phase to the $h_s \!=\! \pm J_3/2$ critical lines of the TSSI XY model is linked to consistently small oscillation amplitudes in the mode decomposition of the LE. Analogous to the ESSH model away from the IP, this can be attributed to the fact that the quasiparticles which control the LE remain fully gapped as one approaches the QPT.
On the contrary, the revival structures which {\em do} appear in the TSSI LEs are associated with large oscillation terms in the mode decomposition of the LE, with amplitudes that peak at wave numbers where nearby quasiparticles have a sizable group velocity. This, in turn, emulates the scenario for the ESSH model at the IP, but now for quenches to special parameter values which do {\em not} define a critical point of a QPT. One should here note that while a QPT may favor large LE oscillation amplitudes \cite{Venuti2010} (however $-$ as transpires from our analysis $-$ only if these are controlled by the quasiparticles which become massless at the QPT), large amplitudes can incidentally appear also {\em within} a quantum phase if this phase supports massless excitations. Provided that these excitations have sizable group velocities, an observable revival structure may then emerge, as evidenced when quenching to the {\em noncritical} $(J_3\! \neq\! 0, h_s\!=\!0)$ line within the type-I spin-liquid phase of the TSSI XY model, cf. Fig. \ref{fig4}.

{\em Summary.} $-$ We have shown that the presence of a quantum phase transition is neither a sufficient nor a necessary condition for observing a revival structure in the Loschmidt echo after a quantum quench. Periodic revivals are preconditioned by a LE controlled by massless quasiparticle modes with a group velocity $v_g \!\gg \!L/t$, where $L$ is the length of the system and $t$ is the observation time. This property may or may not be present at a quantum critical point. The suppression of a critical revival structure is strikingly illustrated away from the isotropic quantum critical point in the extended Su-Schrieffer-Heeger model, introduced in this Letter. Here the revivals are found to be controlled by quasiparticle states which remain gapped at the anisotropic quantum phase transition, implying small oscillation amplitudes in the mode decomposition of the LE. Our findings may call for a revisit of earlier results on revival structures and quantum criticality, and should encourage efforts to identify more reliable nonequilibrium markers of quantum criticality.
%\newpage
\begin{acknowledgments}
{\em Acknowledgments} $-$
H. J. thanks Wen-Long You for valuable discussions. R. Jafari  would like to dedicate this paper to Prof. Y. Sobouti the founder of Institute for Advanced Studies in Basic Sciences.
This research was supported by STINT (Grant No. IG2011-2028) and the Swedish Research Council (Grant No. 621-2014-5972).
\end{acknowledgments}

%\bibliographystyle{prl}
%\bibliography{References}

\vskip 15 cm
\begin{widetext}

\setcounter{figure}{0}
\setcounter{equation}{0}
\setcounter{section}{0}

\newpage

\section{Supplementary material}
\renewcommand\thefigure{S\arabic{figure}}
\renewcommand\theequation{S\arabic{equation}}

\vskip 0.5 cm

In this Supplemental Material we have collected some useful results on the extended Su-Schrieffer-Heeger (ESSH) model, introduced in the accompanying Letter, Ref. \cite{JJ2016}.

\subsection{A. Eigenstates and eigenvalues of the ESSH model}
By Fourier transforming the ESSH Hamiltonian $H$ in Eq. (2) in \cite{JJ2016}, choosing $\mu=0$, and grouping together terms with $k$ and $-k$, $H$ is transformed into a sum of commuting Hamiltonians $H_k$, each describing a different $k$ mode,
\be
\label{eqS1}
H_k= q_{k}c_{k}^{A\dag}c_{-k}^{B\dag} +p_{k}c_{k}^{A\dag}c_{k}^{B}
+ q_{-k}c_{-k}^{A\dag}c_{k}^{B\dag} + p_{-k} c_{-k}^{A\dag}c_{-k}^{B} + \mbox{H. c.}
\ee
Here $p_{k}\!=\!-(w+\tau e^{-ika})$ and $q_{k}\!=\!-(\Delta e^{-i\theta}-\Lambda e^{i(\theta-ka)})$, with $a=1$ the lattice spacing.
We can thus obtain the spectrum of the ESSH model by diagonalizing each Hamiltonian mode $H_k$ in (\ref{eqS1}) independently. This can be done in two ways: Using a generalized Bogoliubov transformation which maps $H_k$ onto the BdG quasiparticle Hamiltonian $H(k)$ in (3) in \cite{JJ2016}  (with the quasiparticle operators $\gamma_k^{\alpha}$ and $\gamma_k^{\alpha \dagger}, \alpha=1,2,3,4$, expressed in terms of the fermion operators in (\ref{eqS1})), or using a basis in which the eigenstates of $H_k$ are obtained as linear combinations of even-parity fermion states \cite{SunChen}. Here we outline the connection between the two approaches.

As $H_k$ in (\ref{eqS1}) conserves the number parity (even or odd number of fermions), it is sufficient to consider the even-parity subspace of the Hilbert space. This subspace is spanned by the eight basis vectors
\begin{align}
\label{eqS2}
|\varphi_{1,k}\rangle&=|0\rangle, &  |\varphi_{2,k}\rangle&=c_{k}^{A\dag}c_{-k}^{A\dag}|0\rangle, &
|\varphi_{3,k}\rangle&=c_{k}^{A\dag}c_{-k}^{B\dag}|0\rangle, &
|\varphi_{4,k}\rangle&=c_{-k}^{A\dag}c_{k}^{B\dag}|0\rangle, \no \\
|\varphi_{5,k}\rangle&=c_{k}^{B\dag}c_{-k}^{B\dag}|0\rangle, &  |\varphi_{6,k}\rangle&=c_{k}^{A\dag}c_{k}^{B\dag}|0\rangle, &
|\varphi_{7,k}\rangle&=c_{-k}^{A\dag}c_{-k}^{B\dag}|0\rangle, &
|\varphi_{8,k}\rangle&=c_{k}^{A\dag}c_{-k}^{A\dag}c_{k}^{B\dag}c_{-k}^{B\dag}|0\rangle.
\end{align}

The eigenstates $|\psi_{m,k}\rangle$ of $H_k$ in this basis can be written as
\be
|\psi_{m,k}\rangle
=\sum_{j=1}^{8}v_{m,k}^{j}|\varphi_{j,k}\rangle,
\ee
where $|\psi_{m,k}\rangle$ is an unnormalized eigenstate of $H_k$
with corresponding eigenvalue $\epsilon_{m,k}\, (m=0,\cdots,7)$, and where
$v_{m,k}^{j}\, (j=1,\cdots,8)$ are functions of the amplitudes of the hopping ($w,\tau$) and pairing terms ($\Delta,\Lambda$), the pairing phases $\pm \theta$,
and the momentum $k$.
Four eigenstates are degenerate with eigenvalues zero ($\epsilon_{2,k}=\epsilon_{3,k}=\epsilon_{4,k}=\epsilon_{5,k}=0$), with the ground state and the first excited state having negative energies
($\epsilon_{0,k}=-\epsilon_{7,k}=(\varepsilon_{k}^{1}+\varepsilon_{k}^{2}), \epsilon_{1,k}=-\epsilon_{6,k}=(\varepsilon_{k}^{1}-\varepsilon_{k}^{2})$ respectively). Here $\varepsilon_{k}^1$ and
$\varepsilon_{k}^2$ are the quasiparticle energies defined after Eq. (3) in \cite{JJ2016}.

Each eigenstate of $H_k$ can be linked to a state in the BdG formalism via their common eigenvalues.
For instance, the ground state $|\psi_{0,k}\rangle$ of $H_k$ is identified with the BdG mode with the corresponding negative energy quasiparticle states filled,
i.e. $|\psi_{0,k}\rangle
= \gamma^{2\dagger}_{k}\gamma^{1\dagger}_{k}|V_k\rangle$, where $|V_k\rangle$ is the $k^{\text th}$ single-fermion mode of the Bogoliubov vacuum. Since the connection between quasiparticle operators and fermion operators is fixed by the Bogoliubov  transformation, we can calculate the Bogoliubov vacuum $|V\rangle = \bigotimes_k |V_k\rangle$
in terms of the eigenstates of $H_k$:
\be
|V\rangle=\bigotimes_{k}\big(\sum_{j=1}^{8}u_{V,k}^{j}|\varphi_{j,k}\rangle\big),
\ee
where $u_{V,k}^{j}$ are functions of the parameters $w,\tau, \Delta,\Lambda$, and $\theta$, and the momentum $k$. The resulting explicit expression is rather unwieldy.
Let us point out that while the BdG formalism is very convenient for obtaining energy eigenvalues, the fermionic even-parity basis is preferable
for numerically computing matrix elements of the time-evolved states, such as those which enter the Loschmidt echo.
%\subsection*{}
%\newpage
\subsection{B. Quasiparticle spectrum}
The BdG quasiparticle spectrum of the ESSH model is plotted
in Fig. \ref{figS1} at (a) the isotropic point (IP) $w=\Delta=\tau=\Lambda=1$ and (b) at the anisotropic point $w=\Delta=2, \tau=\Lambda=1$. The many-particle groundstate of the ESSH Hamiltonian for $\mu=0$ is obtained by filling the two lowest bands, $\{\varepsilon_{k}^{1}\}_{k=-\pi}^{\pi}$ and $\{\varepsilon_{k}^{2}\}_{k=-\pi}^{\pi}$. As seen, at the IP the energy gap between the $\varepsilon_{k}^{1}$ and $\varepsilon_{k}^{4}=-\varepsilon_{k}^{1}$ bands closes at $k=\pi, \theta=\pi/2$ (Fig. \ref{figS1}(a)) while it is nonzero away from the IP (Fig. \ref{figS1}(b)). In contrast, and as required for the existence of the quantum critical line $\theta_c=\pi/2$, the energy gap between the $\varepsilon_{k}^{2}$ and $\varepsilon_{k}^{3}=-\varepsilon_{k}^{2}$ bands
is closed for {\em all} momenta $k$ at $\theta=\pi/2$ for arbitrary values of $\Delta/\Lambda$. One verifies that the groundstate has a $2^{N/2}$-fold degeneracy at the critical line $\theta=\pi/2$ off the IP, with an enlarged degeneracy $2\times 2^{N/2}$ right at the IP.
\vskip 1.0 cm

%$2\times2^{N/2}$-fold degeneracy
%
%
%%%%%%%%%%%%%%%%%%%%%%%  Fig.S1   %%%%%%%%%%%%%%%%%%%%%%%
\begin{figure*}[hb]
\centerline{
\includegraphics[width=0.48\linewidth,height=0.21\linewidth]{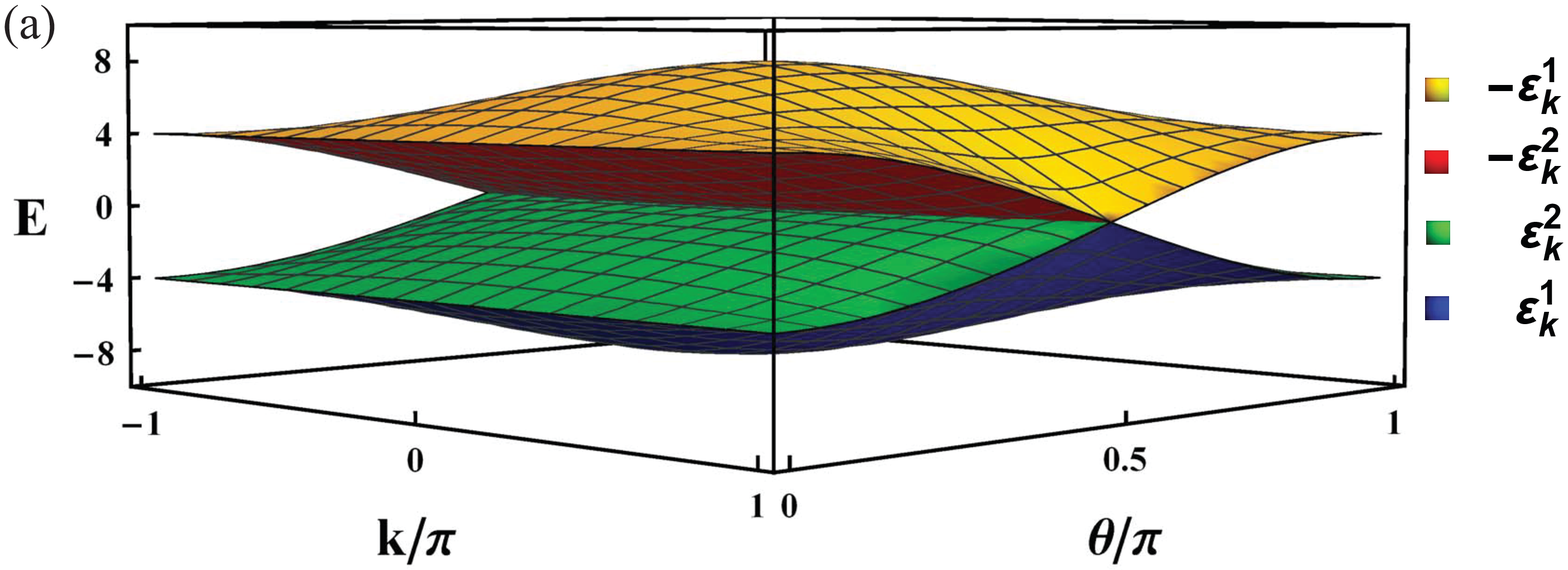}
\includegraphics[width=0.48\linewidth,height=0.21\linewidth]{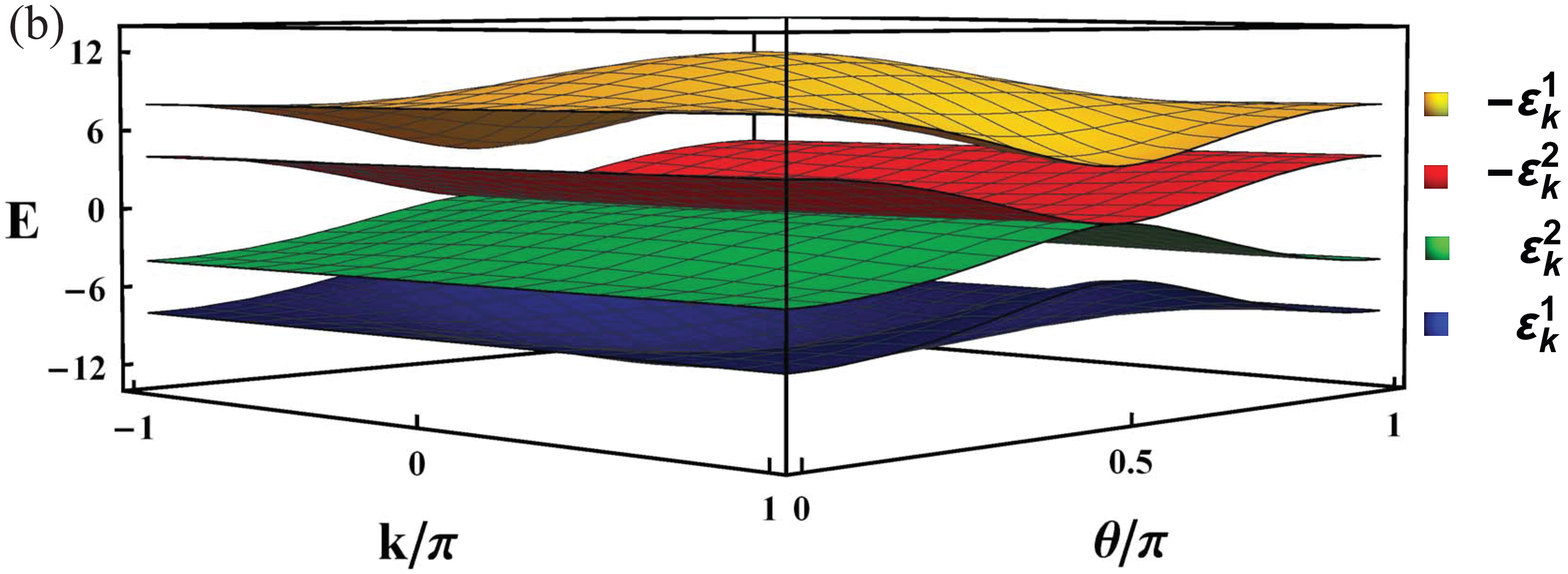}}
\caption{(Color online) BdG quasiparticle spectrum $\{\pm\varepsilon_{k}^{1,2}\}_{k=-\pi}^{\pi}$ for the ESSH model at
(a) the isotropic point (IP) $w=\Delta=\tau=\Lambda=1$, and
(b) at the anisotropic point $w=\Delta=2$, $\tau=\Lambda=1$.}
\label{figS1}
\end{figure*}
%%%%%%%%%%%%%%%%%%%%%%%%%%%%%%%%%%%%%%%%%%%%%%%%%%%%%%%
%
\vskip 1.5 cm

\subsection{C. Loschmidt echo}

The mode decomposition of the Loschmidt echo ${\cal L}({\theta_1},{\theta_2},t)$ in the ESSH model takes the form
\bea
\label{Ldecomp}
{\cal L}({\theta_1},{\theta_2},t)&=& \prod_{{0 \le k \le \pi}}{\cal L}_k({\theta_1},{\theta_2},t),\\
{\cal L}_k(\theta_{1},\theta_{2},t)&=&\Big|\frac{1}{N_{0,k}(\theta_{1})} \langle\psi_{0,k}(\theta_{1})|
e^{-iH(\theta_{2})t}|\psi_{0,k}(\theta_{1})\rangle\Big|^{2}=\Big|\frac{1}{N_{0,k}(\theta_{1})}
\sum_{m=0}^{7}\frac{e^{-i \epsilon_{m,k}(\theta_{2})t}}{N_{m,k}(\theta_{2})}|
\langle\psi_{m,k}(\theta_{2})|\psi_{0,k}(\theta_{1})\rangle|^{2}\Big|^{2} \nonumber \\
&=&\Big|\frac{1}{N_{0,k}(\theta_{1})} \sum_{m=0}^{7}
\frac{e^{-i \epsilon_{m,k}(\theta_{2})t}}{N_{m,k}(\theta_{2})}
\big|\sum_{j=1}^{8}(v_{0,k}^{j}(\theta_{1}))^{\ast}(v_{m,k}^{j}(\theta_{2}))\big|^{2} \Big|^{2}
\eea
where $N_{m,k}(\theta)=|\langle\psi_{m,k}(\theta)|\psi_{m,k}(\theta)\rangle|=\sqrt{\sum_{j=1}^{8}|v_{m,k}^{j}(\theta)|^{2}}$
is the normalization factor of the eigenstate $|\psi_{m,k}(\theta)\rangle$, and where $v_{m,k}^{j}(\theta)$ are
functions of the parameters $\tau, w, \Delta, \Lambda, \theta$ and $\mu$ in the ESSH Hamiltonian, Eq. (2) in \cite{JJ2016}.
For a quench to the critical line $\theta_2 = \theta_c = \pi/2$, the LE reduces to the simple form
\bea
{\cal L}(\theta_{1},\theta_{c},t)= \prod_{0 \le k \le \pi}|1-A_{k}\sin^{2}(\epsilon_{0,k}(\theta_c)t)-B_{k}\sin^{2}(\frac{\epsilon_{0,k}(\theta_c)t}{2})|,
\eea
where $\epsilon_{0,k}(\theta_c)= \varepsilon_{k}^{1}(\theta_c) $ is the ground state energy of $H_k$ at the critical line.
Furthermore, $A_{k}=4(F_{0,k}+F_{1,k})(F_{6,k}+F_{7,k})$ and $B_{k}=4(F_{0,k}+F_{1,k}+F_{6,k}+F_{7,k})(F_{2,k}+F_{3,k}+F_{4,k}+F_{5,k})$, where $F_{m,k}=|\langle\psi_{m,k}(\theta_{c})|\psi_{0,k}(\theta_{1})\rangle|^{2}$ $(m=0,\cdots,7)$.
The oscillation amplitudes $A_k$ and $B_k$ are plotted versus $k$ in Fig. \ref{figA} in \cite{JJ2016}, at the IP and away from the IP. As seen in the figure, the $B_k$-amplitudes at the IP for $k$ approaching the BZ boundary are significantly larger than those away from the IP.

\vskip 1.5 cm
\subsection{D. Fidelity}

By considering the ground state of the system as the initial state, the LE can be interpreted as a dynamical
version of the squared {\em ground-state fidelity} $F(\theta,\theta')$, defined by the overlap between two ground states at
different parameter values $\theta$ and $\theta'$: $F(\theta,\theta')=|\langle \Psi_{0}(\theta)|\Psi_{0}(\theta')\rangle|$.
The ground-state fidelity $F(\theta,\theta \!+\! \delta \theta)$ of the ESSH model can be decomposed as
\bea
F(\theta,\theta \!+\! \delta \theta) &=& \prod_{0 \le k \le \pi}F_{k}(\delta\theta)
\eea
\bea
F_{k}(\delta\theta)=|\langle\psi_{0,k}(\theta)|\psi_{0,k}(\theta\!+\!\delta\theta)\rangle| = \Big|\sum_{j=1}^{8}(v_{0,k}^{j}(\theta))^{\ast}(v_{0,k}^{j}(\theta\!+\!\delta\theta))\Big|,
\eea
with  $v_{0,k}^{j}(\theta)$ functions of the parameters in the ESSH Hamiltonian, Eq. (2) in \cite{JJ2016}. A ground-state fidelity serves as a marker of QPTs \cite{Zanardib2007}, with the QPT at the ESSH critical line $\theta=\pi/2$ signaled by a sharp decay of $F(\theta,\theta \!+\! \delta \theta)$, see Fig. \ref{figS2}. Intriguingly, as seen in the inset
of this figure, the fidelity develops extrema away from the critical line $\theta=\pi/2$, with a local maximum unfolding as one approaches the isotropic point (IP) $\Delta = \Lambda$. We conjecture that this reflects the enhanced groundstate degeneracy at the IP; cf. Sec. B above.

\vskip 1.0 cm
%
%%%%%%%%%%%%%%%%%%%%%%  Fig.S2   %%%%%%%%%%%%%%%%%%%%%%%
\begin{figure}[h]
\centerline{
%\begin{minipage}{0.95\linewidth}
%\small{(a)} $\delta$
%\includegraphics[width=0.95\linewidth]{./figures/fig1.eps}
\includegraphics[width=0.417\linewidth,height=0.25\linewidth]{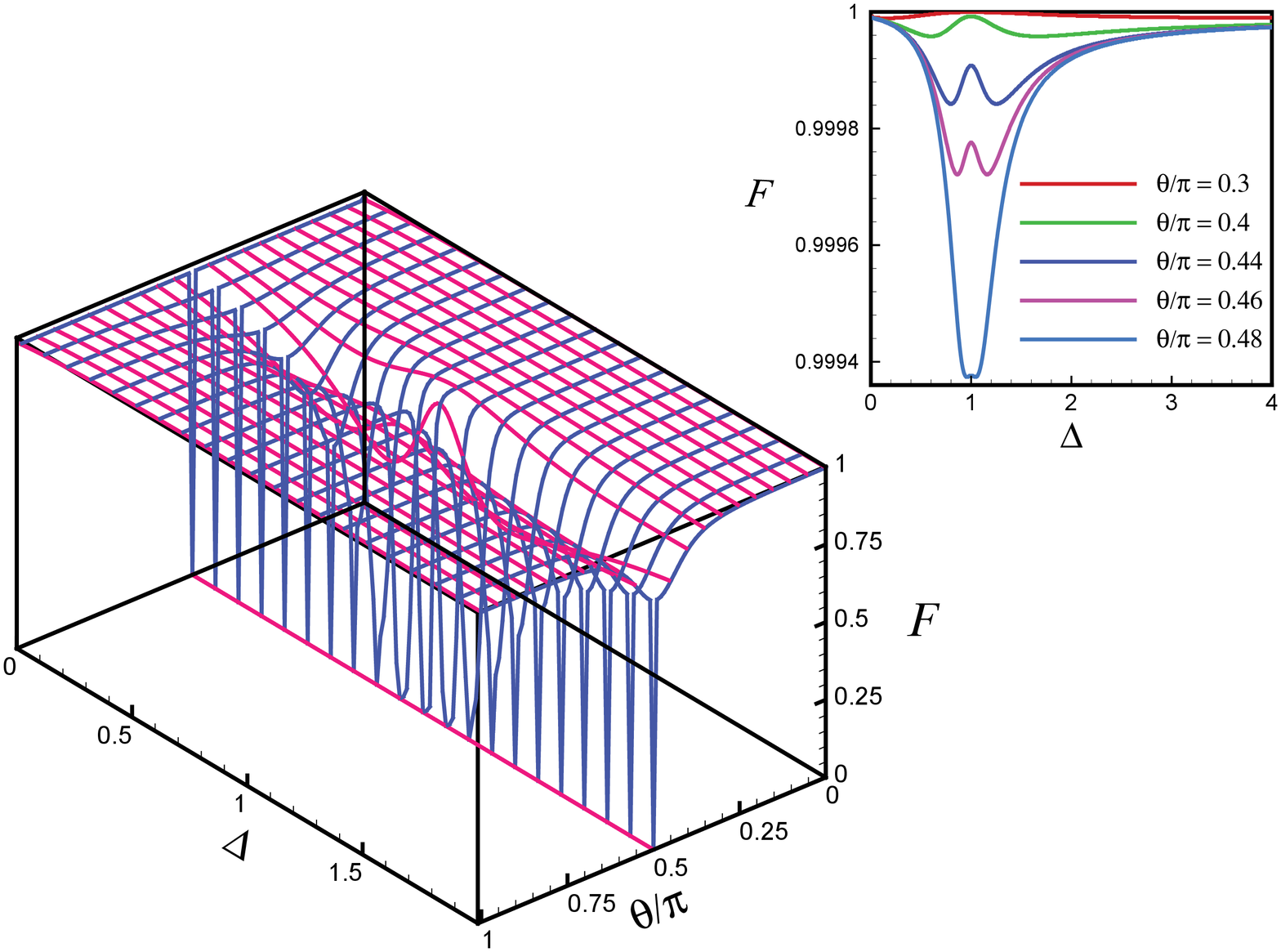}}
%\centering
%\end{minipage}
%\begin{minipage}{0.95\linewidth}
%%\small{(b)} $\delta$
%\includegraphics[width=0.7\linewidth]{./figures/fig2.eps}
%\centering
%\end{minipage}
\caption{ (Color online) Plot of the
ground state fidelity of the ESSH model as a function of the pairing-term amplitude $\Delta$
and phase $\theta$. The other parameters of the model are chosen as
$\tau=\Lambda=1$ and $w=\Lambda$, with $\delta\theta=0.05$ and $N=80$.
Inset: Fidelity of the model versus $\Delta$ for different values of $\theta$ for
$\tau=\Lambda=1, w=\Delta$, $\delta\theta=0.001$, and $N=80$.}
\label{figS2}
\end{figure}
%%%%%%%%%%%%%%%%%%%%%%%%%%%%%%%%%%%%%%%%%%%%%%%%%%
%

\vskip 1.0 cm
\subsection{E. Symmetries and topological phases}
The ESSH Hamiltonian in Eq. (2) in \cite{JJ2016} has particle-hole symmetry, but not time-reversal or chiral symmetry.

To verify this, first note that a particle-hole transformation in the Nambu spinor basis,
defined after Eq. (2) in \cite{JJ2016}, is carried out by the operator
${\cal C}=U_{C}K$ , where $K$ is complex conjugation and $U_{C}=\sigma^{x}\otimes 1$.
One easily checks that ${\cal C}H(k){\cal C}^{-1}=-H(-k)$ where $H(k)$ is the BdG Hamiltonian in Eq. (3) in \cite{JJ2016}.
Turning to the time-reversal operator ${\cal T}$, it is given simply by ${\cal T}=K$ since the fermions in the ESSH model
are spinless. Inspection of $H(k)$ immediately reveals that time-reversal symmetry is broken. The chiral symmetry operator ${\cal S}$ can
by expressed as ${\cal S}={\cal C T}$, and one checks that this symmetry is also broken. The presence of sublattice symmetry, $U_L H(k) U_L^{-1} = - H(k)$, with
$U_L = 1 \otimes \sigma^z$ does not alleviate this fact since in the Nambu spinor basis chiral symmetry does not originate in a lattice substructure.
It follows that the model is in the D symmetry class with a $\mathbb{Z}_2$ topological index \cite{Schnyder2008} .

The $\mathbb{Z}_2$ index is nonzero in the topologically nontrivial phases of the model. These phases can appear when letting
the Hamiltonian parameters take values $w\neq\Delta$ and $\tau\neq\Lambda$ ({\em not} considered in \cite{JJ2016}). For example, for
$w=2, \Delta=\tau=\Lambda=1$, the system is in the Kitaev-like \cite{Kitaevb2001, Wakatsukib2014}
topological phase for $0\leqslant\theta<\pi/3$.
By increasing the pairing
phase $\theta$, the system enters the SSH-like trivial phase \cite{Wang2012} at
$\theta_{c1}=\pi/3$ in which the $\mathbb{Z}_2$ index is zero.
The system once again goes into a Kitaev-like topological phase for
$\theta_{c2} > 2\pi/3$. The revival period of the Loschmidt echo after a quench to the topological phase
transition points $\theta_{c1}=\pi/3$ and $\theta_{c2}=2\pi/3$ is governed by Eq. (5) in \cite{JJ2016}, with $K$ the group velocity of the critical modes \cite{JJ2017}.

\subsection{F. Connection to the general quantum compass model}
The Hamiltonian of the 1D spin-$1/2$ general quantum compass model is given by \cite{Youb2014}
\bea
\label{eqS2}
{\cal H}=
-\sum_{n=1}^{N}
\Big[J_o\tilde{\sigma}_{2n-1}(\theta)\tilde{\sigma}_{2n}(\theta)
+ J_e\tilde{\sigma}_{2n}(-\theta)\tilde{\sigma}_{2n+1}(-\theta) \Big],
\eea
with $J_{e/o}$ exchange amplitudes on even/odd lattice bonds, and
where the pseudo-spin operators $\tilde{\sigma}_n(\pm \theta)$ are
formed by linear combinations of the Pauli matrices $\sigma_{n}^{x}$ and $\sigma_{n}^{y}$:
$\tilde{\sigma}_n(\pm \theta) =\cos(\theta) \sigma_{n}^{x} \pm \sin(\theta) \sigma_{n}^{y}$.
This Hamiltonian can be diagonalized exactly by mapping it onto a free fermion model,
\bea
\label{eqS3}
{\cal H}=-\sum_{n=1}
\Big[(\frac{J_{o}}{4} c^{\dagger}_{2n}c_{2n-1}+\frac{J_{e}}{4} c^{\dagger}_{2n+1}c_{2n}+\mbox{H.c.})
+(\frac{J_{o}}{4} e^{-i\theta} c^{\dagger}_{2n}c^{\dagger}_{2n-1}+\frac{J_{e}}{4} e^{i\theta}c^{\dagger}_{2n+1}c^{\dagger}_{2n}+\mbox{H.c.})\Big],
\eea
using the Jordan-Wigner transformation
\bea
\label{JWT}
\sigma^{+}_{n}=\sigma^{x}_{n} + {\it i}\sigma^{y}_{n}= \prod_{m=1}^{n-1}\left(-\sigma^{z}_{m}\right)c_{n}^{\dagger},~~
\sigma^{-}_{n}=\sigma^{x}_{n} - {\it i}\sigma^{y}_{n}= \prod_{m=1}^{n-1}c_{n}\left(-\sigma^{z}_{m}\right),~~
\sigma^{z}_{n}= 2c_{n}^{\dagger}c_{n}-1.
\eea

By partitioning the chain into bi-atomic
elementary cells and defining two independent fermions at each cell $n$  \cite{Perk1975, Derzhko2009}, $c_{n}^{A}\equiv c_{2n-1}$
and $c_{n}^{B}\equiv c_{2n}$, one can rewrite the Hamiltonian in Eq. (\ref{eqS2}) as
\bea
\label{eqS4}
{\cal H}=-\sum_{n=1}^{N/2}
\Big[
(\frac{J_{o}}{4}c^{A\dagger}_{n}c^{B}_{n}+\frac{J_{e}}{4} c^{A\dagger}_{n+1}c^{B}_{n}+\mbox{H.c.})
+(\frac{J_{o}}{4} e^{-i\theta} c^{A\dagger}_{n}c^{B\dagger}_{n}+\frac{J_{e}}{4} e^{i\theta}c^{A\dagger}_{n+1}c^{B\dagger}_{n}+\mbox{H.c.})\Big].
\eea
Choosing the parameters of the ESSH Hamiltonian, Eq. (2) in \cite{JJ2016}, as $w=\Delta=J_{o}/4$, $\tau=\Lambda=J_{e}/4$, and $\mu=0$, it maps onto the Hamiltonian in Eq. (\ref{eqS4}).
In other words, the ESSH model in this case represents the general quantum compass model. \\
\end{widetext}

%\bibliography{References}
%\bibliography{Ref}
\end{document}